\documentclass[english,9pt,twocolumn,twoside]{IEEEtran}
\usepackage[T1]{fontenc}
\usepackage[latin9]{inputenc}
\usepackage{units}
\usepackage{bm}
\usepackage{amsthm}
\usepackage{amsmath}
\usepackage{amssymb}
\usepackage{graphicx}

\makeatletter

\providecommand{\tabularnewline}{\\}

\theoremstyle{plain}
\newtheorem{thm}{\protect\theoremname}
\theoremstyle{plain}
\newtheorem{prop}[thm]{\protect\propositionname}

\makeatother

\usepackage{babel}
\providecommand{\propositionname}{Proposition}
\providecommand{\theoremname}{Theorem}

\begin{document}

\title{Decision Fusion with Unknown Sensor Detection Probability}

\author{D.~Ciuonzo,~\IEEEmembership{Student Member,~IEEE,} and P.~Salvo~Rossi,~\IEEEmembership{Senior Member,~IEEE}
\thanks{Manuscript received Oct. 23, 2013; revised Nov. 31, 2013;

This work has been partially supported by the ERCIM within the Alain
Bensoussan fellowship programme and by the Faculty of Information
Technology, Mathematics and Electrical Engineering of the Norwegian
University of Science and Technology, Trondheim, Norway, within the
project CAMOS.

The authors are with the Dept. Industrial \& Information Engineering,
Second University of Naples, Aversa (CE), Italy (e-mail: \{domenico.ciuonzo,
salvorossi\}@ieee.org).%
}\vspace{-0.5cm}
}
\maketitle
\begin{abstract}
In this correspondence we study the problem of channel-aware decision
fusion when the sensor detection probability is not known at the decision
fusion center. Several alternatives proposed in the literature are
compared and new fusion rules (namely ``ideal sensors'' and ``locally-optimum
detection'') are proposed, showing attractive performance and linear
complexity. Simulations are provided to compare the performance of
the aforementioned rules.\end{abstract}

\section{Introduction}

\PARstart{D}{ecision} fusion (DF) in wireless sensor networks (WSNs)
attracted huge interest by the scientific community \cite{Varshney1996}.
In some particular cases, assuming that the sensor probability of
detection is higher than the corresponding false-alarm, the uniformly
most powerful test is independent on the local sensor probabilities
\cite{Ciuonzo2013} and thus their knowledge is not needed. However,
it is typically assumed that the sensor performance is known at the
DF center (DFC) \cite{Lei2010,Chen2004,Jiang2005a}. Indeed in the
general case sensor performance is required in order to implement
the optimal fusion rule, namely the likelihood ratio test (LRT). Unluckily,
while the sensor false-alarm can be obtained (since it depends on
the local threshold value and the sensing noise distribution), the
detection probability is generally difficult to acquire, as it depends
on the features of the (unknown) event being observed. 

There are two common approaches tackling the aforementioned problem:
$(i$) employing (sub-optimal) rules which \emph{neglect the whole
sensor performance}, such as the ``diversity'' statistics proposed
in \cite{Lei2010,Chen2004,Ciuonzo2012}; ($ii$) assuming the knowledge
of the local false alarm probabilities and considering the detection
probability as an unknown (deterministic) parameter, thus determining
a \emph{composite hypothesis test}%
\footnote{In the latter case it is assumed that the sensor detection probability
is the same for all the sensors employed (i.e. a homogeneous scenario).%
}. A first remarkable study in the latter direction is found in \cite{Wu2010}
where a fusion rule, obtained along the same lines of a generalized
LRT (GLRT) derivation, has been proposed and shown to have promising
results, i.e. being an affine statistic and outperforming the GLRT
itself in the considered scenarios.

Unluckily, to the best of our knowledge the two approaches have not
been compared yet, and thus it is not immediate whether the sole knowledge
of the sensors false alarm probabilities is a potential benefit in
the design of efficient fusion rules. Also, another (possibly) useful
information is that the sensor detection probability is typically
higher than the corresponding false alarm probability (since each
``informative'' receiver operating characteristic always outperforms
an unbiased coin). We will show that jointly exploiting both information
\emph{can produce} performance gains. 

In this letter we study channel-aware DF when the \emph{false-alarm}
probability of the generic sensor is \emph{known}, while the \emph{detection}
probability is \emph{unknown}. First, we perform (to best of our knowledge,
for the first time) a detailed comparison of existing fusion alternatives,
not\emph{ requiring }knowledge of \emph{sensor detection probability},
based on the approaches ($i$) (i.e. the \emph{counting rule} \cite{Varshney1996})
and ($ii$) (i.e. the rule proposed in \cite{Wu2010}, denoted here
as ``\emph{Wu rule}''). The comparison is strengthened by a theoretical
analysis in the case of a large number of sensors, based on deflection
measures \cite{Picinbono1995}. Also, we derive two novel rules, based
on ``ideal sensors'' assumption (approach ($i$)) \cite{Lei2010,Chen2004,Ciuonzo2013a}
and locally-optimum detection (approach ($ii$)) \cite{Kassam1988}.
For all the considered rules high/low signal-to-noise ratio (SNR)
optimality properties are established in a scenario with identical
sensors and a discussion on complexity and required system knowledge
is reported. Finally, the case of non-identical sensors is considered.

The paper is organized as follows: Sec. \ref{sec:System Model} introduces
the model; in Sec. \ref{sec:Fusion rules} we derive and study the
fusion rules, while in Sec. \ref{sec:Fusion rules nid} we generalize
the analysis to the case of non-identical sensors; in Sec. \ref{sec:Numerical-Results}
we compare the presented rules and confirm the theoretical findings
through simulations; finally in Sec. \ref{sec:Conclusions} we draw
some conclusions; proofs are confined to the Appendix.

\section{System Model\label{sec:System Model}}

The model is described as follows%
\footnote{\emph{Notation} - Lower-case bold letters denote vectors, with $a_{n}$
being the $n$th element of $\bm{a}$; $\left\Vert \bm{a}\right\Vert _{p}$
denotes the $\ell_{p}$-norm of $\bm{a}$; upper-case calligraphic
letters, e.g. $\mathcal{A}$, denote finite sets; $\mathbb{E}\{\cdot\}$,
$\mathrm{var\{\cdot\}}$ and $(\cdot)^{t}$ denote expectation, variance
and transpose, respectively; $P(\cdot)$ and $p(\cdot)$ are used
to denote probability mass functions (pmf) and probability density
functions (pdf), respectively, while $P(\cdot|\cdot)$ and $p(\cdot|\cdot)$
their corresponding conditional counterparts; $\mathcal{N}_{\mathbb{C}}(\mu,\sigma^{2})$
denotes a proper complex-valued Gaussian pdf with mean $\mu$ and
variance $\sigma^{2}$, while $\mathcal{Q}(\cdot)$ is the complementary
cumulative distribution function of a standard normal random variable;
$\mathcal{U}(a,b)$ denotes a uniform pdf with support $[a,b]$; finally
the symbol $\sim$ means \textquotedblleft{}distributed as\textquotedblright{}.%
}. We consider a decentralized binary hypothesis test, where $K$ sensors
are used to discriminate between the hypotheses of the set $\mathcal{H}=\{\mathcal{H}_{0},\mathcal{H}_{1}\}$,
representing the absence ($\mathcal{H}_{0}$) or the presence ($\mathcal{H}_{1}$)
of a specific phenomenon of interest. The a priori probability of
$\mathcal{H}_{i}\in\mathcal{H}$ is denoted $P(\mathcal{H}_{i})$.
The $k$th sensor, $k\in\mathcal{K}\triangleq\{1,2,\ldots,K\}$, takes
a binary decision $d_{k}\in\mathcal{H}$ about the phenomenon on the
basis of its own measurements, which is then mapped to a symbol $b_{k}\in\{0,1\}$;
without loss of generality (w.l.o.g.) we assume that $d_{k}=\mathcal{H}_{i}$
maps into $b_{k}=i$, $i\in\{0,1\}$.

The quality of the $k$th sensor decisions is characterized by the
conditional probabilities $P(b_{k}|\mathcal{H}_{j})$: we denote $P_{D}\triangleq P\left(b_{k}=1|\mathcal{H}_{1}\right)$
and $P_{F}\triangleq P\left(b_{k}=1|\mathcal{H}_{0}\right)$ the probabilities
of detection and false alarm of the $k$th sensor, respectively. Initially,
we assume conditionally independent and identically distributed (i.i.d.)
decisions; this restriction will be relaxed in Sec. \ref{sec:Fusion rules nid}.
Also we assume $P_{D}>P_{F}$, because of the informativeness of the
decision at each sensor. Differently from \cite{Chen2004}, we assume
that\emph{ $P_{F}$ is known }at the DFC\emph{, }but on the other
hand that the true\emph{ $P_{D}$ is unknown}, as studied in \cite{Wu2010}.

The $k$th sensor communicates to the DFC over a dedicated binary
symmetric channel (BSC) and the DFC observes a noisy binary-valued
signal $y_{k}$, that is $y_{k}=b_{k}$ with probability $\left(1-P_{e,k}\right)$
and $y_{k}=\left(1-b_{k}\right)$ with probability $P_{e,k}$, which
we collect as $\bm{y}\triangleq\left[\begin{array}{ccc}
y_{1} & \cdots & y_{K}\end{array}\right]^{t}$. Here $P_{e,k}$ denotes the bit-error probability (BEP) of the $k$th
link%
\footnote{Throughout this letter we make the reasonable assumption $P_{e,k}\leq\frac{1}{2}$.%
}. The BSC model arises when separation between sensing and communication
layers is performed in the design phase (namely a ``\emph{decode-then-fuse}''
approach \cite{Ciuonzo2012}) .

The pmf of $\bm{y}$ is the same under both $\mathcal{H}_{0}$ and
$\mathcal{H}_{1}$, except that the value of the unknown parameter
$P_{1}\triangleq P(b_{k}=1|\mathcal{H})$ is different. After denoting
the pmf with $P(\bm{y};P_{1})$ the test is summarized as:
\begin{equation}
\mathcal{H}_{0}\,:\, P_{1}=P_{F};\qquad\mathcal{H}_{1}\,:\, P_{1}>P_{F};\label{eq: One-sided test}
\end{equation}
which is recognized as a \emph{one-sided} (composite) test \cite{Kay1998}.

\section{Fusion Rules \label{sec:Fusion rules}}

The final decision at the DFC is performed as a test comparing a signal-dependent
fusion rule $\Lambda(\bm{y})$ and a fixed threshold $\gamma$:
\begin{equation}
\Lambda(\bm{y})\begin{array}{c}
{\scriptstyle \hat{\mathcal{H}}=\mathcal{H}}_{1}\\
\gtrless\\
{\scriptstyle \hat{\mathcal{H}}=\mathcal{H}}_{0}
\end{array}\gamma\label{eq:Hypothesis testing}
\end{equation}
where $\mathcal{\hat{\mathcal{H}}}$ denotes the estimated hypothesis.
Hereinafter we propose different fusion rules for the considered problem. 

\textbf{(Clairvoyant) LRT }-\textbf{ }in this case we assume that
also $P_{D}$ is known at the DFC. The explicit expression of the
LRT is given by
\begin{align}
\Lambda_{{\scriptscriptstyle \mathrm{LRT}}} & \triangleq\ln\left[\frac{P(\bm{y};P_{1}=P_{D})}{P(\bm{y};P_{1}=P_{F})}\right]=\sum_{k=1}^{K}\ln\left[\frac{P(y_{k};P_{1}=P_{D})}{P(y_{k};P_{1}=P_{F})}\right]\nonumber \\
 & =\sum_{k=1}^{K}\left\{ y_{k}\ln\left[\frac{\alpha_{k}(P_{D})}{\alpha_{k}(P_{F})}\right]+(1-y_{k})\ln\left[\frac{\beta_{k}(P_{D})}{\beta_{k}(P_{F})}\right]\right\} \label{eq:Clairvoyant_LRT- BSC}
\end{align}
where $\alpha_{k}(P_{1})\triangleq P(y_{k}=1;P_{1})=\left((1-2\, P_{e,k})\cdot P_{1}+P_{e,k}\right)$
and $\beta_{k}(P_{1})\triangleq P(y_{k}=0;P_{1})=\left(1-\alpha_{k}(P_{1})\right)$.
It is apparent that Eq.~(\ref{eq:Clairvoyant_LRT- BSC}) should not
be intended as a realistic element of comparison, but rather as an
optimistic upper bound on the achievable performance (since it makes
use of both $P_{D}$ and $P_{F}$). Differently, in this letter it
is assumed that $P_{e,k}$ can be easily obtained, as in \cite{Chaudhari2012}.

\textbf{Ideal sensors (IS) rule }-\textbf{ }we obtain\emph{ }this
rule by assuming that the sensing phase works ideally, that is $(P_{D},P_{F})=(1,0)$.
This simplifying assumption is exploited in Eq. (\ref{eq:Clairvoyant_LRT- BSC}),
thus leading to:
\begin{equation}
\Lambda_{{\scriptscriptstyle \mathrm{IS}}}\triangleq\sum_{k=1}^{K}(2\, y_{k}-1)\ln\left[\frac{1-P_{e,k}}{P_{e,k}}\right].\label{eq:Ideal sensors- rule}
\end{equation}
The assumption behind Eq. (\ref{eq:Ideal sensors- rule}) is not new:
indeed it was considered in \cite{Lei2010,Chen2004,Ciuonzo2013a}
to derive sub-optimal rules (i.e. the \emph{maximum ratio} and the
\emph{equal gain combiners}) under different communication models.

\textbf{Locally-optimum detection (LOD) rule }- the one-sided nature
of the test considered allows to pursue a LOD-based approach, whose
implicit expression is given by \cite[chap. 6]{Kassam1988,Kay1998}

\begin{equation}
\Lambda_{{\scriptscriptstyle \mathrm{LOD}}}\triangleq\left.\frac{\partial\ln\left[P(\bm{y};P_{1})\right]}{\partial P_{1}}\right|_{P_{1}=P_{F}}\times\left(\sqrt{I(P_{F})}\right)^{-1},\label{eq:LOD-general}
\end{equation}
where $I(P_{1})$ represents the \emph{Fisher information} (FI), that
is:
\begin{equation}
I(P_{1})\triangleq\mathbb{E}\left\{ \left(\frac{\partial\ln\left[P(\bm{y};P_{1})\right]}{\partial P_{1}}\right)^{2}\right\} .
\end{equation}

The explicit form of $\Lambda_{\mathrm{{\scriptscriptstyle LOD}}}$
is shown in Eq. (\ref{eq:LOD-explicit-BSC}) at the top of the next
page; the derivation is given in the Appendix.

\begin{figure*}[t]
\begin{equation}
\Lambda_{{\scriptscriptstyle \mathrm{LOD}}}={\displaystyle \left(\sum_{k=1}^{K}\frac{(1-2\, P_{e,k})\cdot\left[\left(y_{k}-P_{e,k}\right)-(1-2\, P_{e,k})\, P_{F}\right]}{\alpha_{k}(P_{F})\beta_{k}(P_{F})}\right)}\times\left(\sqrt{\sum_{k=1}^{K}\frac{(1-2\, P_{e,k})^{2}}{\alpha_{k}(P_{F})\beta_{k}(P_{F})}}\right)^{-1}\label{eq:LOD-explicit-BSC}
\end{equation}

\hrulefill 
\vspace*{0pt}
\end{figure*}

\textbf{Counting rule (CR) }- this rule is widely used in DF (due
to its simplicity and no requirements on system knowledge) and it
is obtained by assuming that the communication channels are ideal,
i.e.
\begin{equation}
\Lambda_{{\scriptscriptstyle \mathrm{CR}}}\triangleq\sum_{k=1}^{K}y_{k},\label{eq:Counting_rule}
\end{equation}
since $P_{e,k}=0$ entails $\alpha_{k}(P_{1})=P_{1}$ and irrelevant
terms are incorporated in $\gamma$ through Eq. (\ref{eq:Hypothesis testing}). 

\textbf{Wu rule} \cite{Wu2010}\textbf{ }- this rule was proposed
by \emph{Wu et al.} and it was shown to outperform a GLRT rule for
all the scenarios considered. We report only the final result and
omit the details. First an \emph{approximate}%
\footnote{This was derived under a high-SNR assumption \cite{Wu2010}.%
}\emph{ maximum-likelihood }(ML)\emph{ }estimate of $P_{D}$ is obtained
as
\begin{equation}
\hat{P}_{D}\triangleq\frac{1}{K}\sum_{k=1}^{K}\left[(1+2\, P_{e,k})y_{k}-P_{e,k}\right],\label{eq:Pd_estimate_TSP2010}
\end{equation}
then the following statistic is employed:
\begin{equation}
\Lambda_{{\scriptscriptstyle \mathrm{Wu}}}\triangleq(\hat{P}_{D}-P_{F}).\label{eq:detector_TSP2010}
\end{equation}

\textbf{Remark}\label{ Remark}: when $P_{e,k}=P_{e}$ all the rules
are \emph{equivalent}%
\footnote{We use the term \textquotedblleft{}equivalent\textquotedblright{}
to refer to statistics which are equal up to a scaling factor and
an additive term (both independent on $\bm{y}$ and finite), thus
leading to the same performance \cite{Kay1998}.%
}. Thus, when the SNR goes to infinity (i.e. $P_{e,k}\rightarrow0$)
all the rules undergo the same performance. The only exception is
$\Lambda_{\mathrm{\mathrm{{\scriptscriptstyle IS}}}}$, since $\lim_{P_{e,k}\rightarrow0}\Lambda_{\mathrm{{\scriptscriptstyle IS}}}=+\infty$
(such a difference leads to a loss in performance, as shown in Sec.~\ref{sec:Numerical-Results}).
Differently, in the low SNR regime their behaviour is significantly
different, as shown by the following proposition.
\begin{prop}
When the SNR is low at each link, $\Lambda_{{\scriptscriptstyle \mathrm{IS}}}$
and $\Lambda_{{\scriptscriptstyle \mathrm{LOD}}}$ approach $\Lambda_{{\scriptscriptstyle \mathrm{LRT}}}$,
while $\Lambda_{\mathrm{{\scriptscriptstyle Wu}}}$ does not.\label{prop:low-SNR proof}\end{prop}
\begin{IEEEproof}
$\Lambda_{\mathrm{{\scriptscriptstyle IS}}}$ and $\Lambda_{\mathrm{{\scriptscriptstyle LOD}}}$
are \emph{equivalent} to $\sum_{k=1}^{K}\psi(P_{e,k})\, y_{k}$ and
$\sum_{k=1}^{K}\phi(P_{e,k})\, y_{k}$, respectively, where $\psi(P_{e,k})\triangleq\ln\left[\frac{1-P_{e,k}}{P_{e,k}}\right]$
and $\phi(P_{e,k})\triangleq\frac{(1-2\, P_{e,k})}{\alpha_{k}(P_{F})\beta_{k}(P_{F})}$(cf.
Eqs.~(\ref{eq:Ideal sensors- rule}-\ref{eq:LOD-explicit-BSC})).
Also, $\Lambda_{{\scriptscriptstyle \mathrm{LRT}}}=\sum_{k=1}^{K}(\chi(P_{e,k})\, y_{k}+\vartheta(P_{e,k})\,(1-y_{k}))$,
where we have denoted $\chi(P_{e,k})\triangleq\ln\left[\frac{\alpha_{k}(P_{D})}{\alpha_{k}(P_{F})}\right]$
and $\vartheta(P_{e,k})\triangleq\ln\left[\frac{\beta_{k}(P_{D})}{\beta_{k}(P_{F})}\right]$.
When the SNR is small, we can approximate each $\psi(P_{e,k})$, $\phi(P_{e,k})$,
$\chi(P_{e,k})$ and $\vartheta(P_{e,k})$ by a first-order Taylor
series around $P_{e,k}=\frac{1}{2}$. Exploiting these expansions
leads to $\sum_{k=1}^{K}\psi(P_{e,k})\, y_{k}\approx2\sum_{k=1}^{K}(1-2\, P_{e,k})y_{k}$,
$\sum_{k=1}^{K}\phi(P_{e,k})\, y_{k}\approx4\sum_{k=1}^{K}(1-2\, P_{e,k})y_{k}$
and $\Lambda_{\mathrm{{\scriptscriptstyle LRT}}}\approx2\,(P_{D}-P_{F})\sum_{k=1}^{K}\left[(1-2\, P_{e,k})(2\, y_{k}-1)\right]$.
Then, the Taylor-based approximations at low SNR are \emph{all equivalent}
and thus $\Lambda_{{\scriptscriptstyle \mathrm{IS}}}$, $\Lambda_{{\scriptscriptstyle \mathrm{LOD}}}$
and $\Lambda_{{\scriptscriptstyle \mathrm{LRT}}}$ undergo the same
performance. Finally, since $\Lambda_{\mathrm{{\scriptscriptstyle Wu}}}$
is \emph{equivalent} to $\sum_{k=1}^{K}(1+2\, P_{e,k})y_{k}$ (cf.
Eqs. (\ref{eq:Pd_estimate_TSP2010}-\ref{eq:detector_TSP2010})),
at low SNR it poorly approximates $\Lambda_{\mathrm{{\scriptscriptstyle LRT}}}$,
whose Taylor-based approximation is instead \emph{equivalent} to $\sum_{k=1}^{K}(1-2\, P_{e,k})y_{k}$. 
\end{IEEEproof}
It is worth noting that: ($i$) Prop. \ref{prop:low-SNR proof} does
not require $P_{e,k}$ to be equal and that ($ii$) the low-SNR optimality
of $\Lambda_{{\scriptscriptstyle \mathrm{IS}}}$ in Prop. \ref{prop:low-SNR proof}
is coherent with the results shown in \cite{Chen2004,Jiang2005a,Ciuonzo2012}. 

\textbf{Wu rule vs CR deflection comparison:} since all the considered
rules are equivalent to scaled sums of independent Bernoulli random
variables, the pmf $P(\Lambda|\mathcal{H}_{i})$ is intractable \cite{Wu2010}.
Hence we rely on the so-called \emph{deflection measures} \cite{Picinbono1995}\emph{
}$D_{i}\triangleq\frac{\left(\mathbb{E}\{\Lambda|\mathcal{H}_{1}\}-\mathbb{E}\{\Lambda|\mathcal{H}_{0}\}\right)^{2}}{\mathrm{var}\{\Lambda|\mathcal{H}_{i}\}}$
to perform a theoretical comparison between $\Lambda_{{\scriptscriptstyle \mathrm{CR}}}$
and $\Lambda_{{\scriptscriptstyle \mathrm{Wu}}}$. This choice is
justified since, as $K$ grows large, $P(\Lambda|\mathcal{H}_{i})$
converges to a Gaussian pdf (in virtue of the \emph{central limit
theorem} \cite{Papoulis1991}). It can be shown that for CR and Wu
rule the deflections assume the following expressions:
\begin{align}
D_{{\scriptscriptstyle \mathrm{CR}},i}= & \frac{\left(\sum_{k=1}^{K}m_{k}\right)^{2}}{\sum_{k=1}^{K}c_{i,k}},\quad D_{{\scriptscriptstyle \mathrm{Wu}},i}=\frac{\left(\sum_{k=1}^{K}n_{k}m_{k}\right)^{2}}{\sum_{k=1}^{K}n_{k}^{2}c_{i,k}},\label{eq: Deflections CR and Wu}
\end{align}
where $m_{k}\triangleq(1-2P_{e,k})(P_{D}-P_{F})$, $n_{k}\triangleq(1+2P_{e,k})$,
$c_{0,k}\triangleq\alpha_{k}(P_{F})\left(1-\alpha_{k}(P_{F})\right)$
and $c_{1,k}\triangleq\alpha_{k}(P_{D})\left(1-\alpha_{k}(P_{D})\right)$.
W.l.o.g., we assume $P_{e,k}\geq P_{e,k+1}$, which in turn gives
$m_{k}\leq m_{k+1}$, $n_{k}\geq n_{k+1}$ and $c_{i,k}\geq c_{i,k+1}$
(since we assume $P_{e,k}\leq\frac{1}{2}$). Consequently, the \emph{Chebyshev's
sum} \emph{inequalities} \cite{Hardy1988} $\sum_{k=1}^{K}n_{k}m_{k}\leq\frac{1}{K}\left(\sum_{k=1}^{K}m_{k}\right)\left(\sum_{k=1}^{K}n_{k}\right)$
and $\sum_{k=1}^{K}n_{k}^{2}c_{i,k}\geq\frac{1}{K}\left(\sum_{k=1}^{K}c_{i,k}\right)\left(\sum_{k=1}^{K}n_{k}^{2}\right)$
hold, which jointly give:
\begin{equation}
D_{{\scriptscriptstyle \mathrm{Wu}},i}\leq D_{{\scriptscriptstyle \mathrm{Wu}},i}\left(\frac{\sqrt{K}\left\Vert \bm{n}\right\Vert _{2}}{\left\Vert \bm{n}\right\Vert _{1}}\right)^{2}\leq D_{{\scriptscriptstyle \mathrm{CR}},i}
\end{equation}
where $\bm{n}\triangleq\left[\begin{array}{ccc}
n_{1} & \cdots & n_{K}\end{array}\right]^{t}$ and the first inequality arises from the application of \emph{Cauchy-Schwartz
inequality} \cite{Bernstein2009} to $\left\Vert \bm{n}\right\Vert _{1}$.

\begin{figure}
\centering{}\includegraphics[width=0.9\columnwidth]{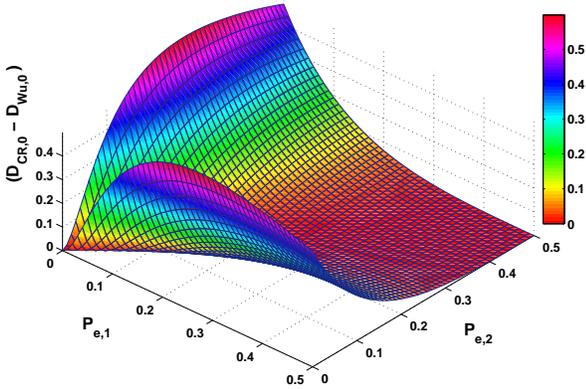}\caption{$\left(D_{{\scriptscriptstyle \mathrm{CR},0}}-D_{{\scriptscriptstyle \mathrm{Wu},0}}\right)$
for $K=2$ sensors as a function of $\{P_{e,1},P_{e,2}\}$, conditionally
i.i.d. decisions $(P_{F},P_{D})=(0.05,0.5)$.\label{fig: D_CR - D_Wu}}
\end{figure}

In Fig. \ref{fig: D_CR - D_Wu} we illustrate $(D_{{\scriptscriptstyle \mathrm{CR}},0}-D_{{\scriptscriptstyle \mathrm{Wu}},0})$
(in a WSN with $K=2$) as a function of $(P_{e,1},P_{e,2})$ in a
scenario with $(P_{F},P_{D})=(0.05,0.5)$. It is confirmed that $D_{{\scriptscriptstyle \mathrm{Wu}},i}$
is always dominated by $D_{{\scriptscriptstyle \mathrm{CR}},i}$ and
that the effect is more pronounced when $P_{e,1}$ and $P_{e,2}$
differ significantly (indeed when $P_{e,1}=P_{e,2}$, $\Lambda_{\mathrm{{\scriptscriptstyle Wu}}}$
is equivalent to $\Lambda_{\mathrm{{\scriptscriptstyle CR}}}$). The
superiority of $\Lambda_{\mathrm{{\scriptscriptstyle CR}}}$ is also
confirmed via the results in Sec.~\ref{sec:Numerical-Results}.

\textbf{Discussion on complexity and system knowledge}: as discussed
in \cite{Wu2010}, $\Lambda_{{\scriptscriptstyle \mathrm{Wu}}}$ being
\emph{affine} in $\bm{y}$ (cf. Eqs. (\ref{eq:Pd_estimate_TSP2010}-\ref{eq:detector_TSP2010}))
is one of the main advantages w.r.t. the GLRT. This feature reduces
the complexity at the DFC and facilitate performance analysis. Since
all the considered alternatives (i.e. $\Lambda_{{\scriptscriptstyle \mathrm{IS}}}$,
$\Lambda_{{\scriptscriptstyle \mathrm{LOD}}}$ and $\Lambda_{{\scriptscriptstyle \mathrm{CR}}}$)
are also affine functions of $\bm{y}$, \emph{they exhibit the same
advantages}. On the other hand, as summarized in Tab.~\ref{tab:System knowledge requirements},
the presented fusion rules have different requirements in terms of
system knowledge. In fact, while $\Lambda_{{\scriptscriptstyle \mathrm{LOD}}}$
and $\Lambda_{{\scriptscriptstyle \mathrm{Wu}}}$ entail the same
requirements (i.e. $P_{F}$ and $P_{e,k}$)), $\Lambda_{{\scriptscriptstyle \mathrm{IS}}}$
only needs $P_{e,k}$. Finally, $\Lambda_{{\scriptscriptstyle \mathrm{CR}}}$
does not require any parameter for its implementation. 

\begin{table}
\begin{centering}
\caption{Comparison of rules w.r.t. system knowledge requirements.\label{tab:System knowledge requirements}}

\par\end{centering}

\smallskip{}

\centering{}%
\begin{tabular}{c||c}
\hline 
\textbf{Fusion rule} &  \textbf{Required parameters}\tabularnewline
\hline 
\hline 
(Clairvoyant) LRT & $P_{D}$, $P_{F}$, $P_{e,k}$ \tabularnewline
\hline 
LOD rule & $P_{F}$, $P_{e,k}$\tabularnewline
\hline 
IS rule & $P_{e,k}$\tabularnewline
\hline 
CR & none\tabularnewline
\hline 
Wu rule \cite{Wu2010} & $P_{F}$, $P_{e,k}$\tabularnewline
\hline 
\end{tabular}
\end{table}

\section{Extension to non-identical sensors scenario\label{sec:Fusion rules nid}}

In this section we generalize the proposed rules to a scenario with
non-identical sensors, i.e. $(P_{D,k},P_{F,k})$, $k\in\mathcal{K}$,
where $P_{F,k}$ is \emph{known} but $P_{D,k}$ is \emph{still unknown}
at the DFC.

\textbf{(Clairvoyant) LRT }- $\Lambda_{\mathrm{{\scriptscriptstyle LRT}}}$
is readily obtained by replacing $\alpha_{k}(P_{D})$ (resp. $\alpha_{k}(P_{F})$)
with $\alpha_{k}(P_{D,k})$ (resp. $\alpha_{k}(P_{F,k})$) in Eq.
(\ref{eq:Clairvoyant_LRT- BSC}). 

\textbf{LOD fusion rule }- the rule is naturally extended to conditionally
independent and non-identically distributed (i.n.i.d.) decisions:
\begin{equation}
\breve{\Lambda}_{\mathrm{{\scriptscriptstyle LOD}}}\triangleq\sum_{k=1}^{K}\left.\frac{\partial\ln\left[P(y_{k};P_{1})\right]}{\partial P_{1}}\right|_{P_{1}=P_{F,k}}\times\left(\sqrt{\mathrm{I}_{k}(P_{F,k})}\right)^{-1}\label{eq:LOD_nid}
\end{equation}

\textbf{CR, IS and Wu fusion rules }- in this scenario $\Lambda_{{\scriptscriptstyle \mathrm{IS}}}$
retains the same form as in Eq. (\ref{eq:Ideal sensors- rule}), while
it is apparent that $\Lambda_{{\scriptscriptstyle \mathrm{CR}}}=\sum_{k=1}^{K}y_{k}$
does not arise from the assumption $P_{e,k}=0$ in $\Lambda_{\mathrm{{\scriptscriptstyle LRT}}}$.
Nonetheless we will still keep $\Lambda_{{\scriptscriptstyle \mathrm{CR}}}$
in the comparison of Sec.~\ref{sec:Numerical-Results}, since it
represents a natural ``$P_{D,k}$-unaware'' alternative. Finally,
we discard Eq. (\ref{eq:detector_TSP2010}) from our comparison, since
the (approximate) ML estimate in Eq.~(\ref{eq:Pd_estimate_TSP2010})
is performed assuming $P_{D,k}=P_{D}$.

\section{Numerical Results\label{sec:Numerical-Results}}

In this section we compare the performance of the proposed rules in
terms of system false alarm and detection probabilities, defined as
\begin{equation}
P_{F_{0}}\triangleq\Pr\{\Lambda>\gamma|\mathcal{H}_{0}\},\qquad P_{D_{0}}\triangleq\Pr\{\Lambda>\gamma|\mathcal{H}_{1}\},
\end{equation}
respectively, where $\Lambda$ is the generic statistic employed at
the DFC.

Similarly as in \cite{Wu2010}, we consider communication over a Rayleigh
fading channel via on-off keying, i.e. $x_{k}=h_{k}b_{k}+w_{k}$,
where $x_{k}\in\mathbb{C}$, $h_{k}\sim\mathcal{N}_{\mathbb{C}}(0,1)$,
$w_{k}\sim\mathcal{N}_{\mathbb{C}}(0,\sigma_{w}^{2})$; $h_{k}$ is
assumed known at the DFC and therefore coherent detection is employed.
Given these assumptions, $P_{e,k}=\mathcal{Q}(\frac{\left|h_{k}\right|}{2\sigma_{w}})$
holds. We define the (individual) communication SNR as the (average
individual) received energy divided by the noise power, that is in
the i.i.d. case 
\begin{equation}
\mathrm{SNR}_{k}\triangleq\frac{\mathbb{E}\{\left|h_{k}b_{k}\right|^{2}\}}{\sigma_{w}^{2}}=\frac{P_{D,k}P(\mathcal{H}_{1})+P_{F,k}P(\mathcal{H}_{0})}{\sigma_{w}^{2}},
\end{equation}
 while in the i.n.i.d. case $\mathrm{SNR}_{\star}\triangleq\mathbb{E}_{(P_{D,k},P_{F,k})}\{\mathrm{SNR}_{k}\}$.
Here we assume $P(\mathcal{H}_{i})=\frac{1}{2}$; the figures are
based on $10^{6}$ Monte Carlo runs. 
\begin{figure}
\centering{}\includegraphics[width=0.9\columnwidth]{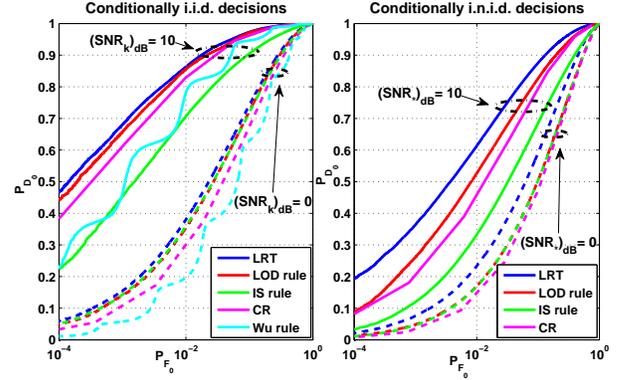}\caption{$P_{D_{0}}$ vs. $P_{F_{0}}$; WSN with $K=10$ and $(\mathrm{SNR}_{k})_{\mathrm{dB}}\in\{0,10\}$
(resp. $(\mathrm{SNR}_{\star})_{\mathrm{dB}}\in\{0,10\}$); $(P_{F,k},P_{D,k})=(0.05,0.5)$
(resp. $(P_{FU},P_{DE})=(0.2,0.6)$) for conditionally i.i.d. (resp.
i.n.i.d.) decisions.\label{fig: ROC iid-inid}}
\end{figure}

In Fig. \ref{fig: ROC iid-inid} we report $P_{D_{0}}$ vs. $P_{F_{0}}$
in a scenario with conditionally i.i.d. and i.n.i.d. decisions, respectively%
\footnote{Note that the concavity of the plots is not apparent, as instead suggested
from the theory \cite{Kay1998}; this is due to the use of a log-linear
scale.%
}. We study a WSN with $K=10$ and local performance equal to $(P_{F,k},P_{D,k})=(0.05,0.5)$
in the i.i.d case while $P_{F,k}\sim\mathcal{U}(0,P_{FU}$), $P_{D,k}=(P_{F,k}+\Delta P)$
and $\Delta P\sim\mathcal{U}(0,P_{DE}$) in the i.n.i.d. case, where
$(P_{FU},P_{DE})=(0.2,0.6)$. We report scenarios with $(\mathrm{SNR}_{k})_{\mathrm{dB}}\in\{0,10\}$
(resp. $(\mathrm{SNR}_{\star})_{\mathrm{dB}}$, where $\mathrm{SNR}_{\star}=\frac{P_{FU}+\nicefrac{P_{DE}}{2}}{2\sigma_{w}^{2}}$
in the i.n.i.d. case). It is apparent that $\Lambda_{{\scriptscriptstyle \mathrm{LOD}}}$
and $\Lambda_{{\scriptscriptstyle \mathrm{IS}}}$ approach $\Lambda_{{\scriptscriptstyle \mathrm{LRT}}}$
at $(\mathrm{SNR}_{k})_{\mathrm{dB}}=0$ in the i.i.d. case (confirming
Prop. \ref{prop:low-SNR proof}), while there is a moderate loss in
the i.n.i.d. case%
\footnote{In fact, it can be verified that Prop. \ref{prop:low-SNR proof} does
not hold in the latter scenario. %
}. However, $\Lambda_{\mathrm{{\scriptscriptstyle \mathrm{IS}}}}$
suffers from significant loss in performance in both cases $(\mathrm{SNR}_{k})_{\mathrm{dB}}=10$
and $(\mathrm{SNR}_{\star})_{\mathrm{dB}}=10$. Also, in the i.i.d.
case $\Lambda_{{\scriptscriptstyle \mathrm{Wu}}}$ is outperformed
by both $\Lambda_{{\scriptscriptstyle \mathrm{CR}}}$ and $\Lambda_{{\scriptscriptstyle \mathrm{LOD}}}$,
the latter being the best choice. Finally, the \emph{oscillating behaviour}
of $\Lambda_{\mathrm{{\scriptscriptstyle Wu}}}$ is explained since
the approximate ML estimate $\hat{P}_{D}$ (cf. Eq. (\ref{eq:Pd_estimate_TSP2010}))
is \emph{not reliable} when the WSN is not of large size. Moreover
the performance of $\hat{P}_{D}$ further degrades at low-medium SNR,
since $\mathbb{E}\{\hat{P}_{D}|\mathcal{H}_{1}\}=\frac{1}{K}\sum_{k=1}^{K}\left((1-4\, P_{e,k}^{2})\cdot P_{D}+2\, P_{e,k}^{2}\right)$,
i.e. when $P_{e,k}^{2}$ is not negligible, the estimator is \emph{biased}
(even if $K$ grows large), as opposed to the exact ML estimate \cite{Kay1993}.

\begin{figure}
\centering{}\includegraphics[width=1\columnwidth]{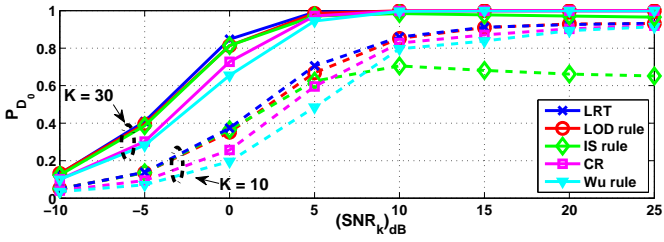}\caption{$P_{D_{0}}$ vs. $\mathrm{(SNR}_{k})_{\mathrm{dB}}$; $P_{F_{0}}=0.01$.
WSN with $K\in\{10,30\}$ sensors; $(P_{F,k},P_{D,k})=(0.05,0.5)$.\label{fig:Pd0 vs SNR}}
\end{figure}

Fig. \ref{fig:Pd0 vs SNR} shows $P_{D_{0}}$ vs. $(\mathrm{SNR}_{k})_{\mathrm{dB}}$,
assuming%
\footnote{In order to keep a fair comparison, we allow for \emph{rule} \emph{randomization}
whenever its discrete nature does not allow to meet the desired $P_{F_{0}}$
exactly. %
} $P_{F_{0}}=0.01$; we simulate a i.i.d. scenario, where $(P_{F,k},P_{D,k})=(0.05,0.5)$
and we report the cases $K\in\{10,30\}$. First, simulations confirm
the theoretical findings in Sec. \ref{sec:Fusion rules}: ($i$) only
$\Lambda_{{\scriptscriptstyle \mathrm{IS}}}$ and $\Lambda_{{\scriptscriptstyle \mathrm{LOD}}}$
approach $\Lambda_{{\scriptscriptstyle \mathrm{LRT}}}$ at low $\mathrm{SNR}$,
while ($ii$) all the considered rules undergo the same performance
as the $\mathrm{SNR}$ increases. The only exception is given by $\Lambda_{{\scriptscriptstyle \mathrm{IS}}}$,
which keeps close to $\Lambda_{{\scriptscriptstyle \mathrm{LRT}}}$
at low-to-moderate $\mathrm{SNR}$ values and exhibits a \emph{unimodal
behaviour}, which is consequence of $\lim_{P_{e,k}\rightarrow0}\Lambda_{\mathrm{{\scriptscriptstyle IS}}}=+\infty$,
as discussed in Sec. \ref{ Remark}. In fact as $P_{e,k}\rightarrow0$,
the possible errors are mainly due to the sensing part; on the other
hand $\Lambda_{{\scriptscriptstyle \mathrm{IS}}}$ assumes a perfect
sensing phase (cf. Eq. (\ref{eq:Ideal sensors- rule})), thus misleadingly
conjecturing that the whole process is error-free. Finally, $\Lambda_{{\scriptscriptstyle \mathrm{LOD}}}$
is close to $\Lambda_{{\scriptscriptstyle \mathrm{LRT}}}$ over the
whole $\mathrm{SNR}$ range considered, while $\Lambda_{{\scriptscriptstyle \mathrm{Wu}}}$
has a significant loss in performance and it is always ``counter-intuitively''
outperformed by $\Lambda_{{\scriptscriptstyle \mathrm{CR}}}$ (with
no requirements on system knowledge). 

\begin{figure}
\centering{}\includegraphics[width=0.9\columnwidth]{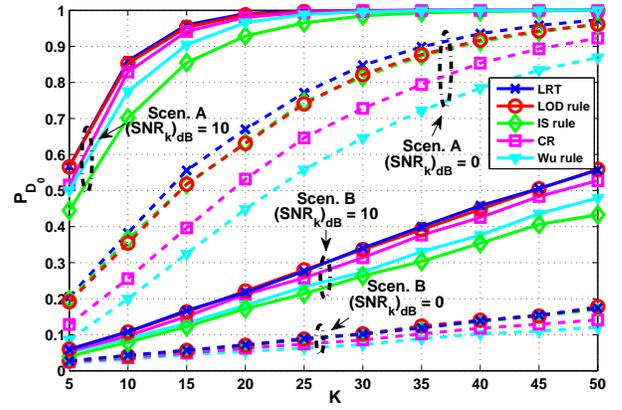}\caption{$P_{D_{0}}$ vs. $K$; $P_{F_{0}}=0.01$. WSN with $(\mathrm{SNR}_{k})_{\mathrm{dB}}\in\{0,10\}$;
$(P_{F,k},P_{D,k})=(0.05,0.5)$ (scen. A) and $(P_{F,k},P_{D,k})=(0.4,0.6)$
(scen. B).\label{fig:Pd0 vs K}}
\end{figure}

Finally, in Fig. \ref{fig:Pd0 vs K} we show $P_{D_{0}}$ vs. $K$,
assuming $P_{F_{0}}=0.01$. We study a i.i.d. setup in the cases $(\mathrm{SNR}_{k})_{\mathrm{dB}}\in\{0,10\}$
(dashed and solid lines, resp.). We analyze the scenarios $(P_{F,k},P_{D,k})=(0.05,0.5)$
(scenario A,\emph{ }as in \cite{Chen2004}) and $(P_{F,k},P_{D,k})=(0.4,0.6)$
(scenario B, as in \cite{Wu2010}). The simulations confirm the performance
improvement given by $\Lambda_{{\scriptscriptstyle \mathrm{LOD}}}$
with respect to $\Lambda_{{\scriptscriptstyle \mathrm{CR}}}$ and
$\Lambda_{{\scriptscriptstyle \mathrm{IS}}}$ (at the expenses of
slightly higher requirements on system knowledge) and the significant
improvement with respect to $\Lambda_{{\scriptscriptstyle \mathrm{Wu}}}$
(the latter being \emph{always} outperformed by $\Lambda_{\mathrm{{\scriptscriptstyle CR}}}$,
even when $K$ is large, as proved in Sec. \ref{sec:Fusion rules}).
For example, in scenario A with $(\mathrm{SNR}_{k})_{\mathrm{dB}}=0$,
$\Lambda_{{\scriptscriptstyle \mathrm{LOD}}}$ achieves $P_{D_{0}}\approx0.8$
with $K\approx30$ sensors as opposed to $K\approx43$ when $\Lambda_{{\scriptscriptstyle \mathrm{Wu}}}$
is employed.

\section{Conclusions\label{sec:Conclusions}}

In this letter we studied DF when the DFC knows the false-alarm probability
of the generic sensor, but does not the detection probability. Wu
rule is always (counter-intuitively, since it makes use of BEPs and
false alarm probabilities) outperformed by the simpler counting rule,
thus does not exploit effectively the required system parameters.
This result is confirmed by a deflection-based analysis, with CR \emph{always}
dominating Wu rule, irrespective of the specific BEPs and local performance
(in the i.i.d case) considered. Differently, the proposed LOD and
IS based rules are appealing in terms of complexity and performance.
LOD rule was shown to be close to the clairvoyant LRT over a realistic
SNR range (thus \emph{effectively} exploiting knowledge of BEPs and
false alarm probabilities), both for conditionally i.i.d. and i.n.i.d.
decisions, as opposed to IS rule (only requiring the BEPs for its
implementation) being close to the LRT only at low-medium SNR. Optimality
of both rules was proved at low SNR in the i.i.d. case, thus motivating
the knowledge of false-alarm probability only at medium SNR in a \emph{homogeneous}
scenario.

\appendix{\label{sec:Appendix- Fisher Information-1}}

We start expressing the log-likelihood $\ln\left[P(\bm{y};P_{1})\right]$
explicitly:
\begin{equation}
\ln\left[P(\bm{y};P_{1})\right]=\sum_{k=1}^{K}\left\{ y_{k}\ln\left[\alpha_{k}(P_{1})\right]+(1-y_{k})\ln\left[\beta_{k}(P_{1})\right]\right\} \label{eq:log-likelihood BSC}
\end{equation}
where $\alpha_{k}(P_{1})$ and $\beta_{k}(P_{1})$ have the same meaning
as in Eq.~(\ref{eq:Clairvoyant_LRT- BSC}). Eq.~(\ref{eq:log-likelihood BSC})
easily provides the numerator in Eq. (\ref{eq:LOD-general}):
\begin{gather}
\frac{\partial\ln\left[P(\bm{y};P_{1})\right]}{\partial P_{1}}=\sum_{k=1}^{K}\frac{\partial\ln\left[P(y_{k};P_{1})\right]}{\partial P_{1}}\nonumber \\
=\sum_{k=1}^{K}\frac{(1-2\, P_{e,k})\cdot\left[\left(y_{k}-P_{e,k}\right)-(1-2P_{e,k})\, P_{1}\right]}{\alpha_{k}(P_{1})\beta_{k}(P_{1})}.\label{eq:der_log-likelihood BSC}
\end{gather}
 On the other hand, we notice that $\mathrm{I}(P_{1})=\sum_{k=1}^{K}\mathrm{I}_{k}(P_{1})$,
where $\mathrm{I}_{k}(P_{1})\triangleq\mathbb{E}\left\{ \left(\frac{\partial\ln\left[P(y_{k};P_{1})\right]}{\partial P_{1}}\right)^{2}\right\} $,
since $y_{k}$ are (conditionally) independent. Hence, we can evaluate
each $\mathrm{I}_{k}(P_{1})$ separately. Considering the explicit
form of $\frac{\partial\ln\left[P(y_{k};P_{1})\right]}{\partial P_{1}}$
in Eq. (\ref{eq:der_log-likelihood BSC}), squaring and taking the
expectation leads to:
\begin{equation}
\mathrm{I}_{k}(P_{1})=(1-2P_{e,k})^{2}\frac{\mathbb{E}\left\{ \left((1-2P_{e,k})P_{1}-(y_{k}-P_{e,k})\right)^{2}\right\} }{\alpha_{k}(P_{1})^{2}\cdot\beta_{k}(P_{1})^{2}}.\label{eq:intermediate_FI-BSC}
\end{equation}

The average in the r.h.s. of Eq. (\ref{eq:intermediate_FI-BSC}) is
given explicitly as follows:
\begin{gather}
\mathbb{E}\left\{ \left((1-2P_{e,k})P_{1}-(y_{k}-P_{e,k})\right)^{2}\right\} =\alpha_{k}(P_{1})\beta_{k}(P_{1}),
\end{gather}
which can be substituted in Eq. (\ref{eq:intermediate_FI-BSC}) to
obtain $\mathrm{I}_{k}(P_{1})$ in closed form. Summing all the (independent)
contributions $\mathrm{I}_{k}(P_{1})$ leads to:
\begin{equation}
\mathrm{I}(P_{1})=\sum_{k=1}^{K}\mathrm{I}_{k}(P_{1})=\sum_{k=1}^{K}\frac{(1-2P_{e,k})^{2}}{\alpha_{k}(P_{1})\cdot\beta_{k}(P_{1})}.\label{eq:FI-BSC}
\end{equation}

Finally substituting Eqs. (\ref{eq:log-likelihood BSC}) and (\ref{eq:FI-BSC})
in Eq. (\ref{eq:LOD-general}) provides Eq. (\ref{eq:LOD-explicit-BSC}).

\bibliographystyle{IEEEtran}
\bibliography{IEEEabrv,sensor_networks}

\end{document}